# Deterministic Task Scheduling in In-Vehicle Networks for Software-Defined Vehicles

Keyvan Aghababaiyan, Baldomero Coll-Perales, Luca Lusvarghi, Javier Gozalvez
Uwicore laboratory, Universidad Miguel Hernandez de Elche, Elche (Alicante), Spain
kaghababaiyan@umh.es, bcoll@umh.es, llusvarghi@umh.es, j.gozalvez@umh.es

*Abstract*— Modern vehicles are embedding increasing levels of automation, connectivity, and intelligence, which require advanced in-vehicle networks and computational platforms to support the dependability and deterministic requirements of critical in-vehicle functions. To this end, the automotive industry is shifting towards software-defined vehicles (SDVs) and zonal E/E architectures with centralized computing nodes. Realizing the full potential of these new architectures requires an efficient management of the in-vehicle's computational workload. In this context, this paper introduces a deterministic task scheduling approach for in-vehicle networks (IVN), and demonstrates that it can better guarantee deterministic service levels than alternative approaches based on the shortest path or the objective to minimize task execution time. Our evaluation also demonstrates that a deterministic task scheduling can satisfactorily support increasing in-vehicle computational workloads and tasks, and achieve a more balanced workload and resource utilization across the IVN. These gains are validated across a variety of IVN topologies, and in hybrid wireless-wired IVN implementations, where a gradual introduction of wireless offers increased in-vehicle connectivity diversity.

*Keywords—Task scheduling, deterministic, SDV, software-defined vehicle, in-vehicle networks, IVN, zonal architecture.*

## I. INTRODUCTION

The rapid automotive evolution has led to an increasing demand for complex and diverse in-vehicle functions, driven by industry trends such as connectivity, electrification, automated driving, and smart mobility. This evolution needs increasingly sophisticated in-vehicle computing capabilities to support features and services with stringent reliability and deterministic service level requirements. Additionally, the computational and operational demands of next-generation automotive systems require evolving from traditional in-vehicle electrical/electronic (E/E) architectures with distributed processing and domain-specific controllers [1]. To address these growing demands, the automotive industry has been shifting towards software-defined vehicles (SDVs), which enable more flexible, configurable, scalable, and upgradable in-vehicle functionalities [2].

The transition to SDVs necessitates significant changes in the in-vehicle network (IVN) and E/E architecture. Traditional architectures, which rely on numerous independent electronic control units (ECUs) to manage sensors and actuators for specific vehicle subsystems, are giving way to zonal IVN architectures with centralized computing [3]. In this new architecture, the vehicle's computational workload is handled by high-performance central compute platforms and zonal controllers, enabling improved coordination between vehicle functions and data fusion, thereby facilitating higher levels of automation and intelligence in decision-making. The zonal IVN architecture with centralized computing also incorporates redundant, diverse, and fault-tolerant elements and links, which are essential for achieving the functional safety levels required in critical systems. However, the efficiency of this IVN architecture depends on the effective management and scheduling of computational tasks, making task scheduling a critical requirement. As vehicle functionalities expand, optimizing workload distribution across processing units is essential to maintain deterministic responsiveness and dependability of critical functions. Traditional static scheduling approaches, where tasks are allocated to predefined computing units or ECUs, may face challenges in this dynamic and demanding environment. The automotive industry advocates for the use of global scheduling with adaptive task allocation approaches to effectively balance workloads across available communication links and computational resources without compromising the execution of critical vehicle functions [1].

In this context, this paper advances the state of the art with the proposal of a deterministic task scheduling approach for zonal in-vehicle E/E architectures with centralized computing. The study demonstrates that a deterministic task scheduling can better guarantee the deterministic service levels of critical in-vehicle functions than alternative state-of-the-art approaches that schedule tasks based on the shortest path [4], [5] or the objective to minimize task execution time [6]. Our evaluation also demonstrates that a deterministic task scheduling can satisfactorily support increasing in-vehicle computational workloads and tasks, and achieve a more balanced workload and resource utilization across the zonal in-vehicle network. We demonstrate that the benefits achieved with a deterministic task scheduling approach are valid across a variety of IVN topologies, ranging from traditional tree-based topologies to mesh topologies with centralized computing based on realistic case studies [4][7]. These benefits are also maintained considering hybrid wireless-wired IVN implementations, where a gradual introduction of wireless offers increased connectivity diversity for linking sensors and actuators to computing units. The results demonstrate that the deterministic task scheduling approach can better adapt to varying operating conditions while enabling efficient resource utilization, thereby preventing resource saturation and enhancing scalability.

The remainder of this paper is organized as follows. Section II presents a comprehensive review of state-of-the-art in-vehicle networks and topologies. Section III introduces the

This work has been partially funded by the European Commission Horizon Europe SNS JU 6G-SHINE (GA 101095738) project, and by MCIN/AEI/10.13039/501100011033 (PID2020-115576RB-I00, PID2023-150308OB-I00).

system model, describing the overall architecture and the key operating conditions for task scheduling within the IVN. Section IV outlines the proposed deterministic task scheduling strategy and state-of-the-art approaches, such as minimum execution time and shortest path task scheduling. Section V details the simulation setup, including the evaluation parameters and scenarios. Section VI evaluates the proposed deterministic task scheduling strategy and compares it with state-of-the-art mechanisms. Finally, Section VII concludes the paper.

## II. STATE-OF-THE-ART IN-VEHICLE NETWORKS AND TOPOLOGIES

Traditional IVN architectures incorporate one ECU for each in-vehicle electronic function with a very specific control task, and a direct interconnection among them. This approach requires new ECUs and interconnections when new sensors or actuators are required. The significant increase in electronic functions has triggered an evolution of IVN architectures for better scalability. A first evolution has been domain-based IVN architectures with several functional domains (infotainment, powertrain, assisted driving, etc.) managed by domain-specific networking technologies and controllers. Despite its benefits, domain-based IVN architectures experience challenges for developing automotive applications that require cross-domain functionality, a need that is growing with vehicle softwarization and the gradual introduction of autonomous driving functions. Zonal IVN architectures have emerged as an alternative to enhance efficiency as vehicle complexity and functionality increase. Zonal IVN architectures group embedded devices and electronics based on physical location rather than logically or per domain. The zonal IVN architecture locally connects sensors and actuators to zonal controllers or ECUs that are physically and strategically distributed through the vehicle. These zonal controllers rely on a high-speed backbone network to connect to each other and to the vehicle's high-performance central computing platform with advanced processing capabilities. A trend in the evolution of zonal IVN architectures is vehicle-centralized computing [3], and the possibility that sensors/actuators may bypass the zonal ECUs and connect directly to the vehicle's central computing platform.

In line with the transition to zonal IVN architectures with centralized computing, this study analyzes four in-vehicle network topologies, depicted in Fig. 1, which are based on realistic case studies from [4][7]. The topologies share a common structure, defining four in-vehicle zones that represent the front-left, front-right, rear-left and rear-right areas of the vehicle. Each zone includes a zone ECU in addition to the sensors and actuators located within that area. The topologies also incorporate a central High-Performance Computing Unit (HPCU). However, they differ in their degree of connectivity. Fig. 1.a represents a conventional tree IVN topology, where sensors and actuators are connected to their respective zonal ECUs, and the zonal ECUs are connected to the central HPCU. Fig. 1.b represents a basic mesh IVN topology which introduces a connectivity backbone between zone ECUs. The ECUs positioned at opposite ends along the diameter are interconnected via the HPCU. Fig. 1.c follows the IVN topology of the Orion Crew Exploration Vehicle (CEV) as utilized in [4], and we refer to it as cross-zone mesh.

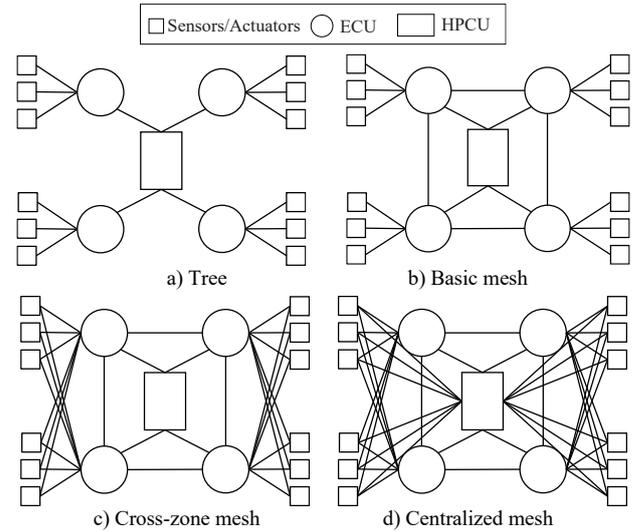

Fig. 1. In-vehicle network topologies.

This topology adds cross-zone connections, providing redundant links between sensors/actuators and nearby zone ECUs. These connections serve as alternative paths for accessing nearby ECUs and the HPCU. Without loss of generality, we consider these cross-zone connections link front and rear sensors/actuators to the zone ECUs located in the opposite area (i.e., left to right and right to left). Finally, Fig. 1.d depicts a centralized mesh topology, which introduces direct links between the sensors/actuators and the HPCU to the cross-zone mesh topology. This provides an alternative path to access the HPCU.

## III. SYSTEM MODEL

The system consists of automotive in-vehicle functions that generate tasks $f_n$, where $n \in \{1, ..., N\}$). Each task $f_n$ is characterized by the tuple $(o_n, c_n, s_n, s'_n, t_n, T_n^{max})$ where: $o_n$ denotes the originating point where tasks can be originated from sensors and actuators ($SNA_{s_m}$), zone ECUs ($zECU_m$), and the HPCU ($H$), where $m \in \{1, ... M\}$ represents the in-vehicle area or zone (M=4) [1], and $s_m \in \{1, ... S_m\}$ is the number of sensors/actuators in the zone $m$. $c_n$ denotes the computing demand of the task, $s_n$ represents the task's size, $s'_n$ is the size of the task after processing, $t_n$ indicates the task generation time, and $T_n^{max}$ defines the task deadline for processing. The processing of tasks is restricted to $zECU_m$ and $H$. We consider that task scheduling schemes (described in Section IV) dynamically assign tasks to computing units within the IVN. When a task $f_n$ is executed on a processing unit different from where it was generated, the processed result with size $s'_n$ must be transmitted back to its source unit. The $T_n^{max}$ of task $f_n$ accounts then for the transmission time to move the task to the assigned processing unit, the processing duration, and the time required to transmit the processed result back to its source unit.

The in-vehicle computing units have different processing speed, denoted as $P_x$ measured in GHz, and a maximum processing capacity $C_x^{max}$ over a time period $T$ with $C_x^{max} = P_x \cdot T$, where $x \in \{z, h\}$ refers to the type of processing unit, i.e., $\{zECU_m, H\}$, respectively. The time required to process a task $f_n$ on a computing unit $x \in \{z, h\}$ is given by:

$$t_p^n = \frac{c_n}{P_x}. \tag{1}$$

We consider that the IVN topologies depicted in Fig. 1 can be fully wired or hybrid wireless-wired. In both cases, we represent with $E$ the set of links between the set of the IVN elements ($SNA_{s_m}$, $zECU_m$, $H$). $d_{ij}$ represents the distance between nodes $i$ and $j$ in the IVN. We consider that nodes that can be reached directly are closer than those that require passing through other nodes.

For wired links, we consider Ethernet-like connections with a data rate $R_w$ and no transmissions errors (i.e., the reliability $\rho$ is equal to 1), ensuring reliable and consistent data transmission. The transmission time over a wired link $w \epsilon E$ can then be computed as:

$$t_c^n = \frac{s_n}{R_W}. \quad (2)$$

In the hybrid wireless-wired IVN scenarios, we restrict the use of wireless connectivity to links between sensors and actuators and the IVN units to which they can connect. The wireless connections therefore depend on the topology (see Fig. 1). Wireless links are prone to transmission errors, and are characterized by a reliability $\rho < 1$. We consider that the wireless links utilize an Orthogonal Frequency Division Multiple Access (OFDMA)-based radio access interface. A dedicated band with bandwidth $BW_m$ is assigned for each of the 4 in-vehicle zones. Additionally, communications between sensors/actuators of all zones and the HPCU in the centralized mesh IVN topology (Fig. 1.d) use a dedicated band of bandwidth $BW_h$. Each $BW_x$ ($x \epsilon \{z,h\}$) is divided into $K_x$ orthogonal resources. Then, the data rate available at any given time for communication resource $k\epsilon\{K_x\}$ in the wireless link $l \epsilon E$ is denoted as $r_l^{(k)}(t)$:

$$r_l^{(k)}(t) = BW_k \cdot log_2(1 + \gamma_l(t))(1 - BER). \quad (3)$$

In (3), $BW_K$ represents the bandwidth of the communication resource $k$, $\gamma_l(t)$ denotes the Signal-to-Interference plus Noise Ratio (SINR) at time $t$ of the wireless link $l$, and BER is the bit error rate, which depends on the modulation and coding scheme employed in the communication resource $k$. To model channel fading effects, we assume a Rayleigh distribution. The total data rate of the link $l$ is calculated as the sum of the data rates for all communication resources $k$ utilized in the link:

$$R_l(t) = \sum_k r_l^{(k)}(t). \quad (4)$$

The transmission time over the wireless link $l$ is then:

$$t_c^n = \frac{s_n}{\bar{R}_l}, \quad (5)$$

where $\bar{R}_l = \frac{1}{\Delta t} \int_{\Delta t} R_l(t) dt$ represents the average of $R_l(t)$ over the time until the end of time slot that $\Delta t$. $\bar{R}_l \geq s_n$. Similarly, the transmission time over wireless communication links for the processed result of a task $f_n$ with size of $s'_n$ can be expressed as $t'^n_c$ and is computed following (5) using $s'_n$ instead of $s_n$.

The total execution time $T_n$ required to complete a task $f_n$ includes the communication time to transmit the task to the processing unit ($t_c^n$), the processing time at the computing unit ($t_p^n$), and the communication time to return the processed result ($t'^n_c$). The total execution time is given by:

$$T_n = t_c^n + t_p^n + t'^n_c. \quad (6)$$

## IV. DETERMINISTIC TASK SCHEDULING

This study proposes a deterministic task scheduling scheme for IVNs. The *Deterministic* scheme prioritizes maximizing the number of tasks completed within their deadlines (i.e., $T_n \leq T_n^{max}$), making it particularly suitable for guaranteeing the timely execution of critical vehicle functions within strict bounded time constraints. The scheme dynamically adjusts task completion times based on varying deadlines, enabling flexible management and balanced workload distribution across the IVN. The objective function is formulated as:

$$min \sum_n \beta\left(\frac{T_n}{T_n^{max}}\right), \quad (7)$$

where $\beta(\xi)$ is a penalty function defined as:

$$\beta(\xi) = \begin{cases} 1 - \prod_{(i,j)\in E_n} x_{ij} \cdot \rho_{ij}, & 0 \leq \xi \leq 1, \\ 1, & \xi > 1, \end{cases} \quad (8)$$

where $x_{ij}$ is a binary decision variable which is equal to 1 if the task is scheduled over the link between node $i$ and node $j$, and $\rho_{ij}$ is the reliability of the link between node $i$ and node $j$. By introducing the reliability of the IVN links in (8), this scheme seeks selecting the most reliable path possible from the multiple available paths when allocating the task from the source to the computing unit. If the selected set of links used to schedule the task has a reliability of 1, no penalty is applied. Conversely, if the combined reliability is less than 1, a penalty (a value between 0 and 1) is introduced, inversely proportional to the overall reliability of the selected links. Eq. (8) also ensures that if a task exceeds its deadline, a penalty of 1 is imposed, discouraging deadline violations.

### A. Constraints

A first binary task scheduling constraint is defined as follows to ensure that task $f_n$ is assigned to a single computing unit and cannot be split among multiple units:

$$\sum_{i=1}^{M+1} a_n^{(i)} = 1, \quad \forall n, \quad (9)$$

where $M+1$ represent the total number of computing units (i.e., $M$ ECUs and 1 HPCU), and $a_n^{(i)}$ is a binary variable equal to 1 if task $f_n$ is allocated to the computing unit $i$.

The second constraint, formulated in eq. (10), is only applicable to wireless links of the hybrid wireless-wired IVN topologies. Following OFDMA principles, this constraint ensures that communication resources from each band are allocated to only one communication link at a time. This prevents transmission collisions and enables interference-free communication between different zones and the HPCU.

$$\sum_{n=1}^{N} b_{l,n,z}^{(k)} = 1, \quad \forall k,l,z, \quad (10)$$

where $b_{l,n,m}^{(k)}$ is a binary variable equal to 1 when communication resource $k$ is allocated to transmit task $f_n$ in link $l$ of band $BW_z$ ($z \epsilon \{m,h\}$).

The third constraint in eq. (11) ensures that the transmission rate for all tasks sharing a link does not exceed the link's maximum achievable data rate.

$$\sum_{n=1}^{N} R_{E,n}(t) \leq R_E(t), \quad \forall l, \quad (11)$$

where $R_{E,n}(t)$ is the data rate of link $E = l \cup w$ for task $f_n$, and $R_E(t)$ is the maximum possible data rate of link $E$.

Finally, the fourth constraint in eq. (12) ensures that the total processing workload of different tasks allocated to a computing unit within a specific time interval does not exceed the unit's maximum processing capacity.

$$\sum_{n=1}^{N} c_n a_n^{(x)} \leq C_x^{max}, \quad \forall x, \quad (12)$$

where $c_n$ denotes the computing demand of the task $f_n$, and $C_x^{max}$ is the maximum processing capacity of unit $x \in \{z, h\}$.

*B. State-of-The-Art Schemes*

The *Deterministic* proposal is compared against three benchmark schemes. The *Baseline* task scheduling scheme follows a traditional static approach, where tasks are allocated to predefined computing units [1]. For the IVN topologies defined in Section II, this means that tasks generated by sensors/actuators are allocated to the ECUs within the same zone, while the ECUs and HPCU process their own tasks.

The *Shortest* task scheduling scheme follows the classic shortest-path approach [4][5] and focuses on minimizing the physical distance between the task source unit and the computing unit. Its objective function is formulated as:

$$\min \sum_{(i,j) \in E} d_{ij} \cdot x_{ij}, \quad (13)$$

where $d_{ij}$ represents the distance between node $i$ and node $j$. $x_{ij}$ is a binary decision variable equal to 1 if the task is scheduled over the link between node $i$ and node $j$, and equal to 0 otherwise. $E = l \cup w$ is the set of IVN links.

The *Minimum* task scheduling scheme follows a common strategy used in task offloading processes [6]. Its objective is to allocate communication resources and computing units to minimize task execution time. This scheme transmits tasks by selecting jointly the fastest available path based on network topology and communication resources and fastest computing unit according to available processing capacity. The optimization function for this strategy is:

$$\min \sum_n T_n, \quad (14)$$

where $T_n$ is defined in (6).

For fairness, the *Shortest* and *Minimum* schemes are defined with the same four constraints as *Deterministic*. Only the first three constrains apply for the *Baseline* scheme since it follows a predefined assignment of the computing units.

## V. EVALUATION SCENARIO

We analyze the impact of task scheduling on the performance of the IVNs following the topologies described in Section II. The IVN consists of 36 sensors/actuators equally distributed in the 4 areas of the vehicle. Each area is controlled by a zone ECU, and central computing is performed in the HPCU. While tasks can be generated by any element of the IVN, only ECUs and the HPCU can handle processing. The processing power of the ECUs and the HPCU is set to 1 GHz and 4 GHz, respectively, based on the existing capabilities of off-the-shelf IVN processing units [8]. Within the vehicle, 70% of tasks are generated by sensors, 15% by ECUs, and 15% by the HPCU. We consider task processing workloads and sizes following the characterization of in-vehicle functions in [9]. In particular, we consider that tasks require between 5 and 15 Mcycles with an average of 10 Mcycles, and their size ranges between 0.5 and 1.5 Mbits, with an average size of 1 Mbits. The size of the processed result for each task is set to 15% of its original size. According to the requirements for in-vehicle functions identified in [7], task deadlines ($T_n^{max}$) are randomly assigned within the 40 to 100 ms range.

When hybrid wireless-wired IVN topologies are considered, the dedicated total bandwidth $BW$ for wireless in-vehicle communication is 100 MHz, divided into five segments: each zone is assigned a bandwidth $BW_m$ of 20 MHz, and the wireless connection to the HPCU has a dedicated bandwidth $BW_h$ of 20 MHz. OFDMA communications are configured with a subcarrier spacing (SCS) of 30 kHz and a time slot duration of 0.5 ms following 3GPP TS 38.211. Based on empirical in-vehicle wireless measurements in [10], we assume that wireless links within the vehicle maintain an average SINR of 30 dB, with channel fading modeled using a Rayleigh distribution. The reliability $\rho$ is considered to randomly vary in the range (0.95–1) for the wireless links between the sensors/actuators and their zonal ECU, and in the range (0.90–1) for the connections to cross-zone ECUs and the HPCU due to the largest distances and presence of blocking elements [10]. The wired links are modeled with an Ethernet-based data rate of 1 Gbps and $\rho=1$.

We implement a genetic algorithm to solve the NP-hard optimization problems of task scheduling as in [7]. The algorithm starts with 1,000 candidate solutions, retaining the top 20% for the next generation while generating the remaining 80% through crossover. Over ten generations, a 20% mutation rate introduces random variations to enhance diversity and prevent premature convergence. This configuration balances performance and computational complexity, achieving near-optimal solutions. The proposed and benchmark schemes are compared for the same number of generations, ensuring a fair comparison by maintaining identical run times.

## VI. RESULTS

We first evaluate the ability of the task scheduling schemes under evaluation to successfully support in-vehicle tasks across different IVN topologies. A task is considered successfully supported if it is executed before its deadline. Fig. 2 depicts the average satisfaction ratio as a function of the number of generated tasks for the fully wired implementation of the four IVN topologies. The satisfaction ratio represents the proportion of tasks completed before their deadlines relative to the total number of tasks. Note that non-satisfied tasks are also completed, but after their deadlines have passed. The results show that the *Baseline* scheme, which relies on pre-assignment of tasks to computing units, can support all generated tasks in scenarios with up to 25 tasks independently of the IVN topology. Its performance significantly degrades with higher workloads. A similar trend is observed for the *Shortest* task scheduling scheme even if it can dynamically schedule tasks across the IVN. This is the case because it does so considering only the physical topology of the IVN to find the shortest path, and does not account for the computing capabilities and workloads of the units or the status of the links in the IVN. The *Minimum* task scheduling scheme does take into account this information to schedule tasks across the IVN to minimize task execution time. This approach outperforms the *Baseline* and *Shortest* schemes across all IVN topologies, and achieves a satisfaction ratio above 95% in scenarios with up to 35 tasks. However, like the *Baseline* scheme, it can only fully execute all tasks within their deadlines in scenarios with up to 25 tasks. On the other hand, Fig. 2 shows that the *Deterministic* scheme can fully satisfy a higher workload and achieves a satisfaction ratio above 95% in scenarios with up to 45 tasks (50 tasks in the centralized mesh). Under this load, the *Deterministic* scheme increases the ratio of satisfied tasks by {26.8%, 27.1%, 8.8%}, {27%, 27.4%, 9%}, {29.7%, 30%, 6.5%} and {30.8%, 30.9%, 6.3%} compared to the {*Baseline, Shortest, Minimum*} schemes for the tree-based, basic mesh, cross-zone mesh and centralized mesh topologies, respectively. These results clearly demonstrate that a deterministic task scheduling approach can better guarantee deterministic service levels and can support increasing in-vehicle computational workloads. In addition, deterministic task scheduling can better leverage advancements in the IVN –such as cross-zonal connections in the cross-zone mesh topology (see Fig. 1.c)– by flexibly managing task completion deadlines to efficiently schedule tasks across the IVN.

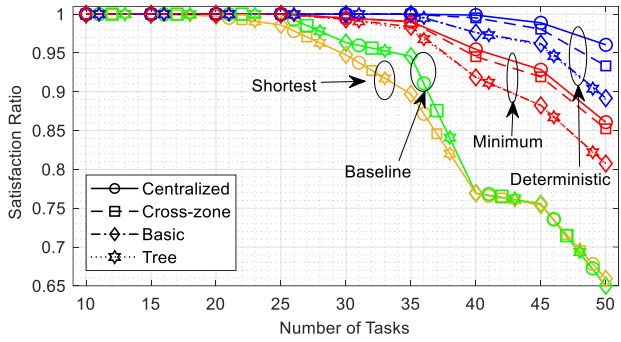

Fig. 2 Task satisfaction ratio for different schemes (wired topologies).

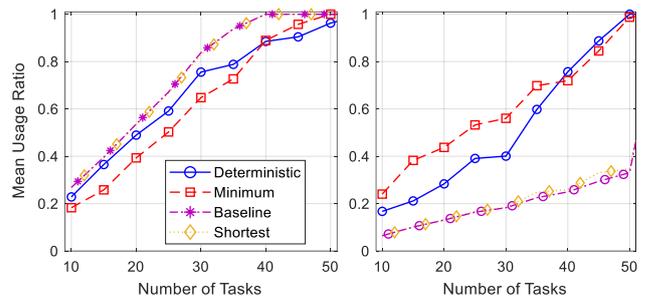

Fig. 4 Usage ratio of computing units (ECUs – left, HPCU – right) in the hybrid wireless-wired centralized mesh IVN topology.

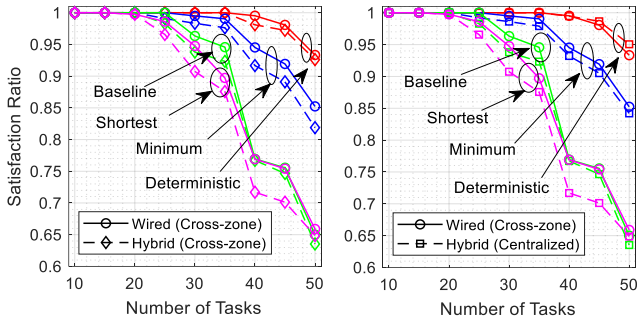

Fig. 3 Task satisfaction ratio for different schemes in wired & hybrid cross-zone (left), and wired cross-zone & hybrid centralized (right).

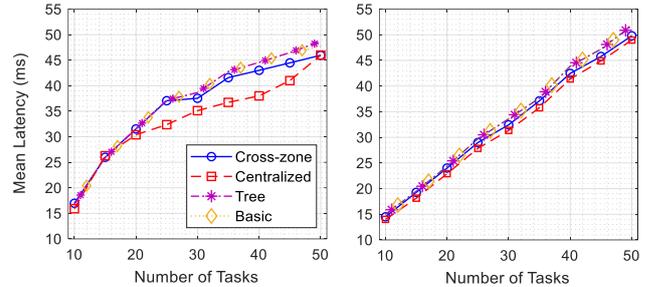

Fig. 5 Latency in different topologies for the *Deterministic* (left) and *Minimum* (right) schemes.

We also analyze the impact of introducing wireless links in the cross-zone mesh IVN topology, specifically in the connections between sensors/actuators and the ECUs. Fig. 3-left compares the satisfaction ratio achieved with the fully wired and hybrid wired-wireless implementations of the topology. The figure shows that all task scheduling schemes experience a reduction in the ratio of satisfied tasks with the introduction of wireless links. However, the reduction is the smallest with the *Deterministic* scheme. For instance, under all considered task loads the reduction it experiences remains below 1.5%, while it increases to 4.3%, 8.6% and 3% for the *Minimum, Shortest* and *Baseline* schemes, respectively. This is because *Deterministic* takes the reliability of the links into account when scheduling tasks to computing units across the IVN. The introduction of wireless links improves the capacity to establish new links within the IVN, and facilitates the flexibility and reconfigurability sought with SDVs. For example, it would be possible to evolve a fully wired cross-zone mesh IVN topology to a hybrid wired-wireless implementation of the centralized mesh topology by adding a wireless link between sensors/actuators and the HPCU (Fig. 1). In this case, Fig. 3-right demonstrates that, with the hybrid centralized mesh IVN topology, the *Deterministic* scheme compensates the performance degradation resulting from the introduction of wireless connections in the hybrid cross-zone mesh IVN topology, and even achieves higher satisfaction ratios compared to the wired cross-zone mesh IVN topology. This is not actually the case for all the other schemes that fail to mitigate the impact of wireless connections, resulting in lower satisfaction ratios with the hybrid centralized mesh IVN topology than with the wired cross-zone mesh IVN topology. The results also show that the *Deterministic* scheme is the only scheme that achieves a satisfaction ratio above 95% in the scenario with 50 tasks in the hybrid centralized mesh IVN topology, outperforming alternative task scheduling schemes by 12.9% to 49.9%.

The higher satisfaction ratios achieved with the *Deterministic* scheme stem from its better scheduling and more balanced workload and resource utilization across the IVN. This is illustrated in Fig. 4, which depicts the average ratio of utilized computing resources of computing units in the hybrid wireless-wired implementation of the centralized mesh topology; similar trends are observed in the other topologies. The left figure shows the average usage ratio of the zone ECUs, while the right figure depicts the usage ratio of the HPCU. Fig. 4 shows that the *Baseline* and *Shortest* schemes saturate the ECUs in the scenarios with 40 tasks or more, while the HPCU experiences a low usage ratio (below 25%) even when the ECUs are saturated. This saturation of the ECUs leads to the drop in the satisfaction ratio shown in Fig. 3 for the *Baseline* and *Shortest* schemes. The *Deterministic* and *Minimum* schemes distribute tasks across different computing units, making better use of the HPCU's high processing power compared to the *Baseline* and *Shortest* schemes. Comparing the *Deterministic* and *Minimum* schemes, Fig. 4 shows that the *Minimum* scheme tends to utilize more the HPCU to minimize the task execution times in scenarios with low to medium task loads. However, under higher task loads, it utilizes the ECUs more than the *Deterministic* scheme. In contrast, the *Deterministic* scheme follows the opposite trend, relying more on the HPCU at higher task loads, which helps avoid bottlenecks in the ECUs by distributing the load more efficiently. This more balanced distribution results in the higher satisfaction ratios shown in Fig. 3 for the Deterministic scheme.

Finally, Fig. 5 compares the latency or total execution time $T_n$ experienced by the Deterministic and Minimum schemes under different IVN topologies. Results are reported for the wired implementation of the tree-based, basic mesh, and cross-zone mesh IVN topologies, and the hybrid wireless-wired implementation of the centralized mesh topology. Fig. 5 shows that *Deterministic* experiences higher latency than *Minimum* in scenarios with low to medium task loads because it prioritizes maximizing the number of tasks completed before their deadlines (Fig. 2-Fig. 3) over minimizing latency. On the other hand, *Deterministic* reduces the latency under higher-load scenarios thanks to its capacity to efficiently adapt the tasks' scheduling to the computing workload, as shown in Fig. 4. Results in Fig. 3 showed that *Deterministic* was the task scheduling approach that could better handle the introduction of wireless links. This is also visible in Fig. 5 that shows that *Deterministic* reduces the latency in the centralized mesh IVN topology by up to 16.3% and 15.6% compared to the tree/basic mesh and cross-zone mesh IVN topologies, while *Minimum* only reduces it by 8.3% and 5.5%, respectively.

## VII. Conclusion

This study has introduced a novel deterministic task scheduling scheme for in-vehicle networks, and has demonstrated its potential to exploit the capabilities of in-vehicle zonal E/E architectures with centralized computing. Our analysis has demonstrated that a deterministic approach to task scheduling can better guarantee deterministic service levels than alternative approaches, and can satisfactorily support increasing in-vehicle computational workloads and tasks. This is achieved thanks to a more balanced workload distribution and resource utilization across the IVN. These trends have been validated across a variety of IVN topologies, and also considering the introduction of wireless connectivity in hybrid IVN topologies.

References

[1] V. Bandur, et al., "Making the Case for Centralized Automotive E/E Architectures", *IEEE Trans. Veh. Technol., vol. 70, no. 2, pp. 1230-1245*, Feb. 2021.

[2] P. Laclau, et al., "Enhancing Automotive User Experience with Dynamic Service Orchestration for Software Defined Vehicles", *IEEE Trans. on Intell. Transp. Syst., vol. 26, no. 1, pp. 824-834*, Jan. 2025.

[3] Robert BoschGmbH, "The next step in E/E architectures", Aug. 2023. Accessed: Mar. 2025. [Online]. Available: https://www.bosch-mobility.com/en/mobility-topics/ee-architecture/

[4] A. Berisa, et al., "AVB-aware Routing and Scheduling for Critical Traffic in Time-sensitive Networks with Preemption", *Proc. ACM 30th RTNS, pp. 207-2018*, Paris (France), 7-8 June 2022.

[5] B. Xu, et al., "A Joint Routing and Time-Slot Scheduling Load Balancing Algorithm for In-Vehicle TSN", *IEEE Trans. Consum. Electron.* (early access on IEEE Xplore since Feb. 2025).

[6] J. Cai et al., "Multitask multi objective deep reinforcement learning-based task offloading method for industrial Internet of Things", *IEEE Internet Things J., vol. 10, no. 2, pp. 1848–1859*, Sep. 2023.

[7] S.D. McLean, et al., "Configuring ADAS platforms for automotive applications using metaheuristics", *Front. Robot. AI, vol. 8*, Jan. 2022.

[8] Advantech, "ECU-4784," Accessed: Mar. 2025. [Online]. Available: https://www.advantech.com/en-us/products/1-369nwl/ecu-4784/mod_18553282-e8f5-4b32-a64b-1083f7182d36.

[9] Y. Deng, et al., "Multi-hop task routing in vehicle-assisted collaborative edge computing", *IEEE Trans. Veh. Technol.*, vol. 73, no. 2, pp. 2444–2455, 2023.

[10] E.A. Vitucci, et al. "6G-SHINE D2.3: Radio propagation characteristics for in-X subnetworks", Dec. 2024.